\RequirePackage{fix-cm}
\documentclass[twocolumn,epjc3]{svjour3}  
\smartqed  
\RequirePackage{graphicx}
\RequirePackage{amsmath}
\usepackage{lineno}
\usepackage{wasysym}

\RequirePackage[colorlinks,citecolor=blue,urlcolor=blue,linkcolor=blue]{hyperref}
\journalname{Eur. Phys. J. C}
\begin{document}


\title{Double-beta decay investigation with highly pure enriched $^{82}$Se for the LUCIFER experiment}

\subtitle{}
\author{
J.~W.~Beeman\thanksref{BERK} \and
 F.~Bellini\thanksref{ROMA1,INFN-RM} \and
 P.~Benetti\thanksref{PV,INFN-PV} \and
 L.~Cardani\thanksref{ROMA1,PRINCE} \and
 N.~Casali\thanksref{ROMA1,INFN-RM} \and
 D.~Chiesa\thanksref{MIB,INFN-MIB} \and
 M.~Clemenza\thanksref{MIB,INFN-MIB} \and
 I.~Dafinei\thanksref{INFN-RM} \and
 S.~Di~Domizio\thanksref{INFN-GE,GE} \and
 F.~Ferroni\thanksref{ROMA1,INFN-RM} \and
 L.~Gironi\thanksref{MIB,INFN-MIB} \and
 A. Giuliani\thanksref{CSNSM} \and
 C.~Gotti\thanksref{INFN-MIB} \and
 M.~Laubenstein\thanksref{LNGS} \and
 M.~Maino\thanksref{MIB,INFN-MIB} \and
 S.~Nagorny\thanksref{GSSI} \and
 S.~Nisi\thanksref{LNGS} \and
 C.~Nones\thanksref{CEA} \and
 F.~Orio\thanksref{INFN-RM} \and
 L.~Pagnanini\thanksref{GSSI} \and
 L.~Pattavina\thanksref{LNGS,corrauthor} \and
 G.~Pessina\thanksref{INFN-MIB} \and
 G.~Piperno\thanksref{ROMA1,INFN-RM} \and
 S.~Pirro\thanksref{LNGS} \and
 E.~Previtali\thanksref{INFN-MIB} \and
 C.~Rusconi\thanksref{INFN-MIB} \and
 K.~Sch\"{a}ffner\thanksref{LNGS} \and
 C.~Tomei\thanksref{INFN-RM} \and
 M.~Vignati\thanksref{INFN-RM} 
}

\thankstext{corrauthor}{e-mail: luca.pattavina@lngs.infn.it}
                                
\institute{
Lawrence Berkeley National Laboratory, Berkeley, California 94720 - USA\label{BERK} \and
 Dipartimento di Fisica - Sapienza Universit\`{a} di Roma, Roma I-00185  - Italy\label{ROMA1} \and
 INFN - Sezione di Roma, Roma I-00185  - Italy\label{INFN-RM} \and
 Dipartimento di Chimica, Universit\`{a} di Pavia, Pavia I-27100 - Italy\label{PV} \and
 INFN - Sezione di Pavia, Pavia I-27100  - Italy\label{INFN-PV} \and
 Physics Department, Princeton University, Princeton, NJ 08544 - USA\label{PRINCE} \and
 INFN - Laboratori Nazionali del Gran Sasso, Assergi (L'Aquila) I-67100 - Italy\label{LNGS} \and
 Dipartimento di Fisica, Universit\`{a} di Milano-Bicocca, Milano I-20126 - Italy\label{MIB} \and
 INFN - Sezione di Milano Bicocca, Milano I-20126 - Italy\label{INFN-MIB} \and
 INFN - Sezione di Genova, Genova I-16146  - Italy\label{INFN-GE} \and
 Dipartimento di Fisica, Universit\`{a} di Genova, Genova I-16126 - Italy\label{GE} \and
 Centre de Spectrometri\'{e} de Masse, Orsay F-91405 - France\label{CSNSM} \and
 Gran Sasso Science Institute, L'Aquila I-67100 - Italy\label{GSSI} \and
 CEA, Irfu, SPP Centre de Saclay, Gif-sur-Yvette F-91191 - France\label{CEA}
}

\date{Received: date / Accepted: date}

\maketitle

\begin{abstract}
The LUCIFER project aims at deploying the first array of enriched scintillating bolometers for the investigation of neutrinoless double-beta decay of $^{82}$Se. The matrix which embeds the source is an array of ZnSe crystals, where enriched $^{82}$Se is used as decay isotope. The radiopurity of the initial components employed for manufacturing crystals, that can be operated as bolometers, is crucial for achieving a null background level in the region of interest for double-beta decay investigations. In this work, we evaluated the radioactive content in 2.5~kg of 96.3\% enriched $^{82}$Se metal, measured with a high-purity germanium detector at the Gran Sasso deep underground laboratory. The limits on internal contaminations of primordial decay chain elements of $^{232}$Th, $^{238}$U and $^{235}$U are respectively: $<$61~$\mu$Bq/kg, $< $110~$\mu$Bq/kg and $<$74~$\mu$Bq/kg at 90\% C.L.. The extremely low-background conditions in which the measurement was carried out and the high radiopurity of the $^{82}$Se allowed us to establish the most stringent lower limits on the half-lives of double-beta decay of $^{82}$Se to 0$^+_1$, 2$^+_2$ and 2$^+_1$ excited states of $^{82}$Kr of 3.4$\cdot$10$^{22}$~y, 1.3$\cdot$10$^{22}$~y and 1.0$\cdot$10$^{22}$~y, respectively, with a 90\% C.L..
\end{abstract}

\section{Introduction}
The observation of neutrinoless double-beta decay (0$\nu\beta\beta$) would demonstrate lepton number violation, and at the same time it would necessarily imply that the neutrinos have a Majorana character. Under the assumption that 0$\nu\beta\beta$ decay is induced by the exchange of light Majorana neutrinos, the partial width of the decay is proportional to the square of the effective neutrino Majorana mass. A measurement of the half-life of this process would supply information to fundamental open questions in Particle Physics: which is the mass hierarchy of neutrinos? Which is their absolute mass scale?\newline

The Standard Model counterpart of the 0$\nu\beta\beta$ decay, is the 2$\nu\beta\beta$ decay. It is the rarest nuclear weak process experimentally observed in dozen of nuclei with half-lives in the range of 10$^{18}$-10$^{22}$~y~\cite{Value_DBD2v}.\newline  

\begin{figure}[h!]
\centering
\includegraphics[width=0.5\textwidth]{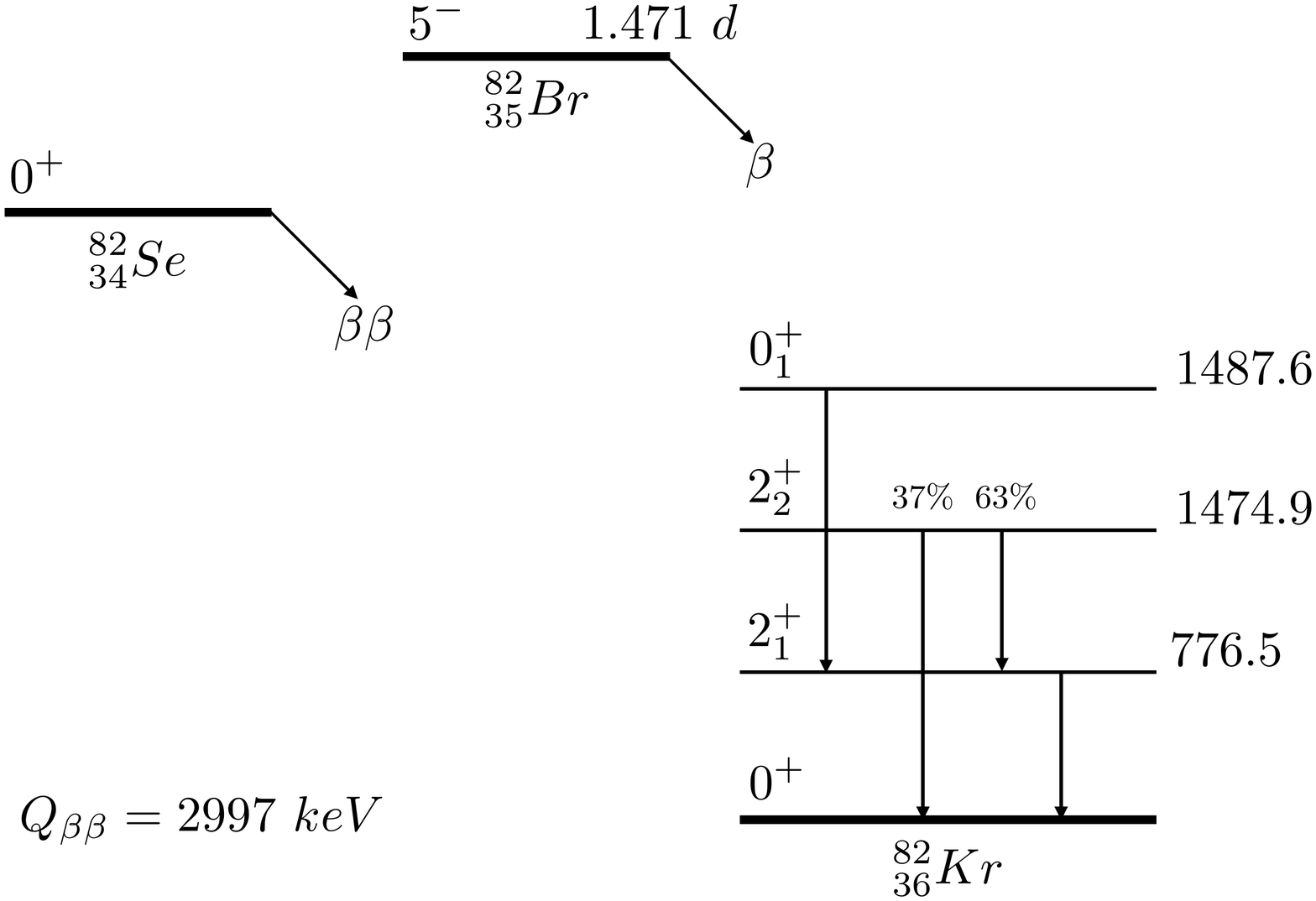}
\caption{\label{fig:Scheme}$^{82}$Se decay scheme. The energies of the excited levels of $^{82}$Kr are reported in keV.}
\end{figure}

Double-beta ($\beta\beta$) decay can occur through different channels other than the ground state to ground state transitions. In Fig.\ref{fig:Scheme} the scheme for $\beta\beta$  $^{82}$Se decay is shown. The transition into different excited levels of the daughter nucleus can also take place, if energetically allowed. The study of $\beta\beta$ decay into excited states can supply futrher information about the nuclear matrix elements that describe 0$\nu\beta\beta$ decay; specifically, a more detailed description of the nuclear model and structure could be gained~\cite{Dhiman_DBD}.\newline

In 1990, the first publication on the potential to observe 2$\nu\beta\beta$ decay to excited levels of daughter nuclei was reported~\cite{Barabash_1st_ex}. Therein an estimation of the expected half-lives for 2$\nu\beta\beta$ decay to the first 0$^+_1$ level for some nuclei was provided. Given the large transition energy involved in such decays - few MeV - the detection seemed to be accessible to the former technology.\newline

Two neutrino double-beta decay to the first 2$^+$ excited level, according to the nuclear matrix elements computed by Haxton and Stephenson~\cite{Haxton_DBD_calc}, seemed to be experimentally inaccessible to detection, due to the smaller phase-space factor compared to the 0$^+$ level. However, later on in the early nineties, J.~Suhonen and O.~Civitarese demonstrated~\cite{Suhonen_exc} that this suppression was not as large as calculated in~\cite{Haxton_DBD_calc} in 1984, and that the observation of these transitions was within reach. In fact, in~\cite{Dhiman_DBD} the half-life for the 2$\nu\beta\beta$ decay of $^{82}$Se to the 2$^+_1$ excited level of $^{82}$Kr was computed to be 5.71$\cdot$10$^{21}$~y.\newline

In order to enhance the experimental sensitivity to the investigation of rare decays it is mandatory to reduce any possible background source. In the $\beta\beta$ decay context, a material selection is needed to single out the $\beta\beta$ decay source with the lowest content of radioactive impurities, which may mimic the signal under investigation. Among the various techniques for material screening a very common one is $\gamma$-spectroscopy using high-purity germanium (HP-Ge) detectors, since they ensure a rather low background level, gain stability and high energy resolution. Such detectors match exactly the needs for a high sensitivity investigation of $\beta\beta$ decay on excited levels of daughter nuclei, where single $\gamma$ quanta or cascades are produced during the de-excitation. In literature, half-life measurements of nuclides which undergo $\beta\beta$ decay on excited levels are mostly performed by means of $\gamma$ spectroscopy~\cite{Tutti_exc,Mo100,Nd150,Gerda_exc}. \newline

In this work, we report on the analysis of the radiopurity level of the $\beta\beta$ decay source for the LUCIFER experiment~\cite{LUCIFER_exp}, namely 96.3\% enriched $^{82}$Se metal, performed at the Gran Sasso Underground Laboratory (LNGS) of INFN, sited in Italy. Thanks to the high purity of the sample, new results on the experimental investigation of $\beta\beta$ decay of $^{82}$Se to the excited states 0$_1^+$, 2$_1^+$ and 2$_2^+$ of $^{82}$Kr are presented.

\section{The LUCIFER project}

In order to detect the elusive signal rate of 0$\nu\beta\beta$ there are two main requirements that must be fulfilled: employing a large mass source and operating the detector in almost zero-background conditions. Present experiments are already designed to investigate the 0$\nu\beta\beta$ decay in ton-scale sources, and a further increase in the source mass seems not to be within the reach of the present technology. Acceptable background rates in the region of interest are of the order of 1-10~counts/y/ton if the goal is just to approach or touch the inverted hierarchy region of neutrino masses, whereas at least one order of magnitude lower values are needed to fully explore it~\cite{CUORE_IHE}. This is the goal of CUPID~\cite{CUPID_1}, a proposed future tonne-scale bolometric neutrinoless double-beta decay experiment to probe the Majorana nature of neutrinos in the inverted hierarchy region of the neutrino mass~\cite{CUPID_2}.\newline

Among the available detection techniques, bolometry is one of the most promising ones. Bolometers ensure high detection efficiency~\cite{MC_bolometer} - the detector embeds the decay source (detector=source) - excellent energy resolution~\cite{CUORE0} and, for luminescent bolometers, also the identification of the nature of the interacting particle~\cite{Bobin,LiMoNe,ZnMoO4_big,TeO2_cherenkov,pbwo4}, an excellent tool for background suppression. Furthermore, this technique shows the unique feature that detectors can be grown out of a wide choice of materials, allowing to probe different isotopes with the same technique.\newline

LUCIFER (Low Underground Cryogenic Installation For Elusive Rates), being one of the first CUPID demonstrators, aims at deploying the first close to background-free small-scale bolometric experiment for the investigation of $^{82}$Se 0$\nu\beta\beta$ decay~\cite{LUCIFER_exp}. Enriched scintillating bolometer of Zn$^{82}$Se will be used for a highly sensitive search for this decay.\newline

The selection of high-purity Zn and $^{82}$Se is one of the most critical point that will undoubtedly affect the sensitivity of the final experiment. In fact, the starting materials for the Zn$^{82}$Se synthesis and the crystal growth must show extremely low concentration of impurities, both radioactive and chemical.

\subsection{Background sources at Q$_{\beta\beta}$}

Selenium is an attractive isotope for the investigation of 0$\nu\beta\beta$ decay, especially for its Q$_{\beta\beta}$-value, which is 2997.9$\pm$0.3~keV~\cite{Se82_Q}. Since it lies above the most intense natural $\gamma$-line at 2.6~MeV, the $\beta/\gamma$ background in the region of interest (ROI) is strongly reduced. Most of the background at this energy in bolometers is expected to come from $\alpha$ induced surface contaminations~\cite{sticking}. Thanks to a highly efficient particle discrimination, taking advantage of the heat-light read-out, such kind of background can be rejected in scintillating bolometers~\cite{Bobin,LiMoNe,ZnMoO4_big,pbwo4}.\newline

Given our present knowledge~\cite{LUCIFER_exp}, the main background sources for the LUCIFER experiment will be: i) high energy $\gamma$ emission from $^{214}$Bi - up to 3.2~MeV - and ii) pile-up events from $^{208}$Tl $\gamma$ cascade. Even if the background induced by these sources can be suppressed, as discussed in~\cite{first_ZnSe}, still it can not be completely neglected. The two previously mentioned nuclides are produced in the decay chains, following the decay of primordial $^{238}$U and $^{232}$Th. A thorough material selection is mandatory in order to reduce as much as possible their concentration in the Zn$^{82}$Se crystals. In order to allow an almost zero-background investigation~\cite{ZnSe_characterization,Laura_Thesis}, $^{238}$U and $^{232}$Th contaminations inside the final crystal must be at the level of tens of $\mu$Bq/kg or better.

\subsection{Chemical impurities}

When selecting the starting material for crystal growth, it is crucial that the raw materials exhibit high purity grades. This issue becomes even more relevant when dealing with bolometers. Point defects caused by impurities in the starting materials may act as traps both for phonons, involved in the bolometric signal development, and for free charge carriers, involved in the scintillation process. The achievement of high purity raw materials becomes more difficult while handling enriched selenium, given the complexity of the enrichment production process requiring large quantities of chemical reagents.\newline

Elements of the VI and VII groups of the periodic table (e.g. V, Cr, Mn, Fe, Co, Ni) are defined as critical impurities for the scintillation performance of Zn$^{82}$Se crystals~\cite{ZnSe_production,ZnSe_scint}. Moreover, ferromagnetic and paramagnetic impurities can spoil the bolometric properties of the detector, since they add a contribution to the overall heat capacity of the system~\cite{Heat_capacity_bolo}. In fact, at low temperatures, low heat capacities are desirable because they will result in large thermal signals and thus higher energy resolutions.

\section{Selenium enrichment}
The entire enrichment process for the production of 15~kg of $^{82}$Se for the LUCIFER experiment was carried out at URENCO, Stable Isotope Group in Almelo, in the Netherlands. In order to be able to achieve the required experimental sensitivity, the use of enriched Se is mandatory, given the relatively low natural isotopic abundance of 8.73\%~\cite{Isotopic_abundance}. The enrichment is performed through a well established procedure: centrifugal enrichment. A dedicated line of centrifuges was employed for the enrichment of gaseous SeF$_6$, that was fully separated from the one used for the $^{235}$U enrichment. The technical details, i.e. the number of centrifuges in the cascade, the throughput and other specifications are property of URENCO. In order to prevent U, Th and other nuclides contaminations in the final product, a flushing of the entire centrifuge cascade was performed using fluorine, and providing a satisfactory cleaning of the system.\newline

The production of enriched Se consists of 3 main steps:
\begin{enumerate}
\item[1-] procurement of SeF$_6$ gas;
\item[2-] centrifugal enrichment of $^{82}$SeF$_6$;
\item[3-] conversion of $^{82}$SeF$_6$ gas to $^{82}$Se metal.
\end{enumerate}

The starting material is natural SeF$_6$ gas, purchased by URENCO from an external supplier, which is fed into a cascade of centrifuges. The outcome of the process is $^{82}$SeF$_6$ enriched gas at a $\geq$ 95\% level. The subsequent step is the conversion from gas to metal, which is performed through a series of chemical reactions~\cite{Benetti_proceeding}. The final product is enriched $^{82}$Se metal.\newline

In order to minimize the risk of losses and of exposure to cosmic rays, thus to reduce cosmogenic activation, the enriched $^{82}$Se was divided into 6 batches, separately delivered to the LNGS by ground transportation and stored underground. Furthermore, special care was devoted to prevent any re-contamination of the isotope~\cite{sticking}, for this reason each batch was packed in double vacuum sealed polyethylene bags before shipping.\newline

For the LUCIFER experiment, 15~kg of enriched selenium were provided in form of small beads. Each gas-to-metal conversion yields about 200~g, for the production of the final enriched metal, 70 different processes were carried out. Each conversion was supplied with a set of witness samples for monitoring the purity and the enrichment level of selenium.\newline

The purity grade of the reagents and the yield of the transformation affect the quality of the final product, for this reason the process is recognized to be most critical for the experiment. Thanks to the constructive collaboration with the producer all reagents employed for the chemical conversion of selenium were subjected to assay both from a chemical and radioactive point of view, using an Inductively Coupled Plasma Mass Spectrometer (ICP-MS) and $\gamma$-spectroscopy with HP-Ge. The contaminations of radioactive elements - U/Th primordial decay chains - and chemical - V, Cr, Mn, Fe, Co and Ni - were measured to be at a level of 10$^{-7}$~g/g . These values are within the requirements for the production of enriched selenium with an overall chemical purity better than 99.8\% on trace metal base. All material screenings were performed at LNGS.\newline 


\begin{table*}[htdp]
\begin{center}
\caption{Isotopic abundance of natural (recommended values~\cite{Isotopic_abundance}) and enriched selenium. The values for the enriched isotope are averaged over 70 different gas-to-metal conversion processes, weighted for their relative mass.} 
\begin{tabular}{ccccccc}
\hline\noalign{\smallskip}
 & $^{74}$Se& $^{76}$Se& $^{77}$Se& $^{78}$Se& $^{80}$Se& $^{82}$Se\\ 
\noalign{\smallskip}\hline\noalign{\smallskip}
Natural Se  [\%] & 0.87 & 9.36 &7.63 & 23.78 &49.61 & 8.73\\
\noalign{\smallskip}\hline\noalign{\smallskip}
Enriched Se  [\%] & $<$0.01 & $<$0.01 &$<$0.01 &$<$0.01 & 3.67$\pm$0.14 & 96.33$\pm$0.31\\  
\noalign{\smallskip}\hline
\end{tabular}
\label{tab:enrichment} 
\end{center}
\end{table*}

\begin{table*}[htdp]
\begin{center}
\caption{Concentration of chemical impurities inside the 15~kg of enriched $^{82}$Se. For each batch we list the gas-to-metal conversion process which showed the highest level of contaminations. The measurements are carried out by means of an Inductively Coupled Plasma Mass Spectrometer at the Gran Sasso laboratory. Limits are computed at 90\% C.L..} 
\begin{tabular}{ccccccccccc}
\hline\noalign{\smallskip}
    \# & Na & S & V & Cr & Mn & Fe & Ni & Co & Th & U \\
              & [ppb] & [ppb] & [ppb] & [ppb] & [ppb] & [ppb] & [ppb] & [ppb] & [ppb] & [ppb]\\
    \noalign{\smallskip}\hline\noalign{\smallskip}
    1 & $<$1000 & 130000 & $<$40 & $<$100 & $<$10 & 110 & 64 & $<$4.0 & $<$0.1 & $<$0.1 \\
    \noalign{\smallskip}\hline\noalign{\smallskip}
    2 & 3300  & 185000 & $<$90 & $<$100 & $<$20 & $<$500 & $<$100 & $<$10 & $<$1.0 & $<$2.0 \\
    \noalign{\smallskip}\hline\noalign{\smallskip}
    3 & $<$2100  & 191000 &$<$4.1 & 32.4 & $<$41.5 & 91.9 &$<$102.5 & $<$2.1 & $<$10.38 & $<$10.38 \\
    \noalign{\smallskip}\hline\noalign{\smallskip}
    4 & $<$1900  & 257000 &  $<$9.89 & 13.3 & $<$36.2 & 590.2 & $<$97.6 & $<$2.0 & $<$0.98 & $<$0.59\\
    \noalign{\smallskip}\hline\noalign{\smallskip}
    5 & 3200  & 236000 & $<$9.4 & 35.8 & $<$37.4 & 697.7 &143.1 & $<$9.4 & $<$0.94 & $<$0.94\\
    \noalign{\smallskip}\hline\noalign{\smallskip}
    6 & $<$2000 & 249000 & $<$10.0 & $<$50.1 & $<$40.1 & 276.4 & 37.7 & $<$10.0 & $<$0.2 & $<$0.2\\ 
\noalign{\smallskip}\hline
\end{tabular}
\label{tab:ICPMS} 
\end{center}
\end{table*}

In Table~\ref{tab:enrichment} the abundance of the different selenium isotopes after the enrichment process, measured with ICP-MS are reported. The values reported in the table are the average over the 70 conversion processes weighted for their relative mass, while the uncertainty refers to the spread in the enrichment. As it is shown in the table, an enrichment above 95\% on $^{82}$Se is ensured on all the samples, in some cases an enrichment level up to 97.7\% was achieved. For the sake of comparison, we report also the recommended natural occurring values~\cite{Isotopic_abundance}.

\section{Chemical and radiopurity assay}

The analysis of chemical impurities inside the 6 different deliveries of $^{82}$Se was performed using ICP-MS. The results are shown in Table~\ref{tab:ICPMS}, where for each batch we show the impurities of the sample which exhibits the highest concentration. Together with the previously mentioned elements, we investigated other impurities whose concentration is induced by the use of chemical reagents for the gas-to-metal conversion. Sodium and sulphur are the elements mostly involved in this chemical conversion~\cite{ZnSe_production}, where they are employed in the form of salt like: NaI or Na$_2$SO$_3$. The concentration of these is not relevant for the final crystal production.\newline

The investigation of the internal radioactive contamination of enriched Se metal for the LUCIFER experiment was carried out with $\gamma$-spectroscopy using a HP-Ge detector located in the underground Laboratory of LNGS (depth of ~3600 m w.e.~\cite{LNGS_depth}), which ensures muon flux suppression by a factor of ~10$^6$ compared to sea level. The detector, GeMPI-4~\cite{GeMPI}, is a p-type germanium crystal provided by Canberra company, with a volume of about 400~cm$^3$. The crystal size was optimized for high counting efficiency in Marinelli type geometry. Details on the experimental set-up and detector performance can be found in~\cite{GeMPI}.\newline

Through $\gamma$-spectroscopy, background information is inferred for the most intense lines of the primordial decay chains $^{238}$U/$^{235}$U, $^{232}$Th and $^{40}$K, but also on elements with antropogenic (e.g. $^{137}$Cs) and cosmogenic ($^{60}$Co, $^{75}$Se) origins. The overall background counting rate of the detector is 5.6~c/keV/kg/y over a large energy range [200,2700]~keV. The FWHM energy resolution of the detector ranges from 1.5~keV at 609~keV to 2.2~keV at 1460~keV, while at higher energies, like 2615~keV, is 3.2~keV. \newline
\begin{figure}[h!]
\centering
\includegraphics[width=0.5\textwidth]{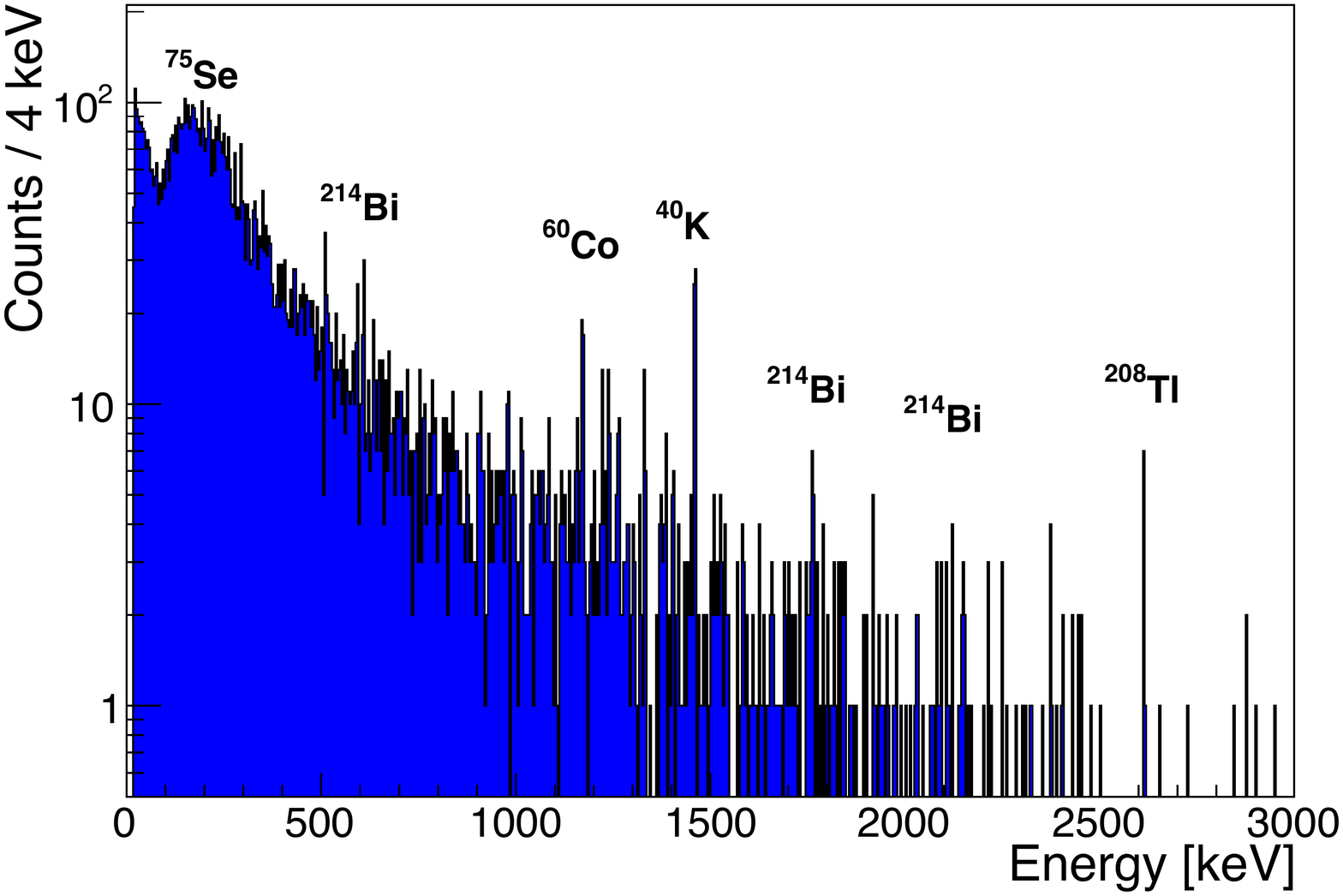}
\caption{\label{fig:spectrumSe82}Gamma energy spectrum acquired over 75~days with 2.5~kg of 96.3\% enriched $^{82}$Se metal beads.}
\end{figure}
Selenium beads were poured inside a 1.46~l volume polypropylene container. A cylindrical Marinelli type beaker was chosen: the detector look at the whole lateral and top surfaces of the sample, whose net mass was 2500.5~g.

\section{Results and limits on contaminations}

Data were acquired over 1798.2~hours, the final energy spectrum is shown in Fig.~\ref{fig:spectrumSe82}. Few lines are visible in the energy spectrum, these are produced by radioactive impurities inside the sample and in the HP-Ge detector itself.\newline

The activity of radioactive nuclides inside the sample is computed following the same standard procedure described in~\cite{LYS}. In the evaluation of the contaminant concentrations we take into account the detector background as well as the detection efficiencies for the different gamma lines, which are calculated with Monte Carlo simulations based on the Geant4 code~\cite{Geant4}.\newline

The internal contaminations of the sample, as well as their concentrations, are listed in Table~\ref{tab:contaminations}. No evidence of nuclides from the natural chains of $^{235}$U/$^{238}$U and $^{232}$Th are detected, in agreement with the ICP-MS measurements (see Table~\ref{tab:ICPMS}). We report the limits of detectability for long-lived $\beta$/$\gamma$-emitters of the decay chains. Limits are set at an extremely low level - hundreds of $\mu$Bq/kg, and they are competitive with values reported on other raw materials used for the crystal growth of $\beta\beta$ decay detectors~\cite{TeO2_production,Amore_production,GERDA}.\newline

Limits on the concentration of other commonly observed nuclides are also shown, specifically for $^{40}$K and $^{60}$Co.

In the sample, $^{75}$Se is the only nuclide having a concentration at a detectable level, 110$\pm$40~$\mu$Bq/kg. This isotope is not naturally occurring and it is mainly produced in significant amount by cosmogenic neutron spallation on $^{76}$Se. The reaction $^{76}$Se($n,2n$)$^{75}$Se has a rather large neutron interaction cross section: 979$\pm$90 mb for 16~MeV neutrons~\cite{Se_activation}. According to~\cite{Se_activation}, among the fast neutron induced reactions on selenium isotopes, $^{75}$Se, with an half-life of 119.8~days, is the only long-lived nuclide produced. $^{75}$Se decays through an electron capture and has a rather small Q-value of 863.6~keV. This makes the isotope not dangerous for $^{82}$Se $\beta\beta$ decay investigations. Moreover the measurement was started in October 2014, thus a strong reduction in the $^{75}$Se activity is expected before the crystal production, that will be finalized in March 2016.

\section{$^{82}$Se double-beta decay to $^{82}$Kr excited levels}
The purity of the enriched selenium and the high performance of the HP-Ge detector allowed us to carry out a highly sensitive investigation of $^{82}$Se $\beta\beta$ decay to the excited levels of $^{82}$Kr. Specifically, we look for the $\gamma$ signatures produced by the de-excitation of the nuclei that undergo a $\beta\beta$ decay to an excited level.\newline

The technique adopted for this investigation does not allow us to distinguish between 0$\nu\beta\beta$ and 2$\nu\beta\beta$ decay of $^{82}$Se. In fact the experimental efficiency for the detection of the two emitted electrons is insignificant, but on the other hand the set-up is sensitive to the $\gamma$ emission from the source. The study of the $\beta\beta$ transition is mainly focused on the first three excited levels of $^{82}$Kr: 0$_1^+$, 2$_1^+$ and 2$_2^+$. The decay scheme for the $^{82}$Se $\beta\beta$ decay is shown in Table~\ref{tab:epsilon}, where the four $\gamma$ emissions under investigation are listed.\newline
We can parametrize the decay scheme as follows:
\begin{eqnarray}
N_a &=& \epsilon_a \cdot \xi \cdot \Gamma_{0^+_1} \; \rightarrow 711~\text{keV}\;\\
N_b &=& \epsilon_b \cdot \xi \cdot BR_b \cdot \Gamma_{2^+_2} \; \rightarrow 1474.9~\text{keV}\;\\
N_c &=& \epsilon_c \cdot \xi \cdot BR_c \cdot \Gamma_{2^+_2} \; \rightarrow 698.4~\text{keV}\;\\
N_d &=& \epsilon_d \cdot \xi \cdot ( \Gamma_{0^+_1} + \Gamma_{2^+_1} + BR_c \cdot \Gamma_{2^+_2} ) \; \rightarrow 776.5~\text{keV}\; \label{eqn:4}
\end{eqnarray}
where $N_x$ ($x=a,b,c,d$) represents the observed number of events for the 4 different $\gamma$'s, $\epsilon_x$ the total detection efficiency of the process, $\xi$ the total exposure (180.40~kg$\cdot$days of $^{82}$Se) and $\Gamma_y$ ($y={0^+_1},{2^+_1},{2^+_2}$) the partial width of the decays. $BR$ represents the branching ratio of the decay which is 37\% for the transition $2^+_2 \rightarrow 0^+$ and 67\% for the $2^+_2 \rightarrow 2^+_1$ (see Fig.\ref{fig:Scheme}). Special care is also addressed to the 776.5~keV line, because it is produced through three different channels, as described by Eq.~\ref{eqn:4}.\newline

The detection efficiencies for the four different $\gamma$ emissions, derived via simulations, are listed in Table~\ref{tab:epsilon}.
\begin{table}[h]
\caption{Detection efficiencies for the 4 different nuclear transitions on the $^{82}$Kr excited levels.} 
\begin{center}
\begin{tabular}{ccccc}
\hline\noalign{\smallskip}
& Transition & Energy [keV] & $\epsilon$ [\%] \\ 
\noalign{\smallskip}\hline\noalign{\smallskip}
a) & $0^+_1 \rightarrow 2^+_1$ & 711 & 3.2 \\
\noalign{\smallskip}\hline\noalign{\smallskip}
b) & $2^+_2 \rightarrow 0^+$ & 1474.9 & 0.3 \\
\noalign{\smallskip}\hline\noalign{\smallskip}
c) & $2^+_2 \rightarrow 2^+_1$ & 698.4 & 2.9 \\
\noalign{\smallskip}\hline\noalign{\smallskip}
d) & $2^+_1 \rightarrow 0^+_1$ & 776.5 & 3.1 \\
\noalign{\smallskip}\hline
\end{tabular}
\label{tab:epsilon} 
\end{center}
\end{table}
In our final calculation we neglect the $2^+_2 \rightarrow 0^+$ transition, given the relatively small detection efficiency.\newline

We simultaneously extract the decay rate for the three excited levels of $^{82}$Kr from the entire data set (180.40~kg$\cdot$days of $^{82}$Se) performing a global Binned Extended Maximum Likelihood fit in the energy range between 650 and 850~keV~(ROI). The probability density function ($p.d.f.$) has four components, a flat background modelling the multi-scatter events produced by high energy $\gamma$'s (e.g. $^{40}$K,  $^{208}$Tl and $^{214}$Bi) and three gaussian functions, with fixed central value ($\mu$) and width ($\sigma$), for the (a), (c) and (d) transitions:
\begin{eqnarray*}
p.d.f. = & \frac{N_{bkg}}{N_{total}}\cdot flat(E) + \\
         + & \frac{N_{a}}{N_{total}}\cdot Gaussian(E,\mu_a,\sigma) + \\
         + & \frac{N_{c}}{N_{total}}\cdot Gaussian(E,\mu_c,\sigma) + \\
         + & \frac{N_{d}}{N_{total}}\cdot Gaussian(E,\mu_d,\sigma).
\end{eqnarray*}
The energy resolution of the detector, which is represented by $\sigma$, is assumed to be constant over the narrow ROI, its value is 0.61~keV. This is evaluated on the 609~keV peak, the closest visible peak to the ROI, and it is produced by natural $^{214}$Bi contaminations in the experimental set-up. The flat background assumption is verified by comparing the counting rate in the ROI and in adjacent regions at higher and lower energies.\newline

\begin{figure}[h]
\centering
\includegraphics[width=0.5\textwidth]{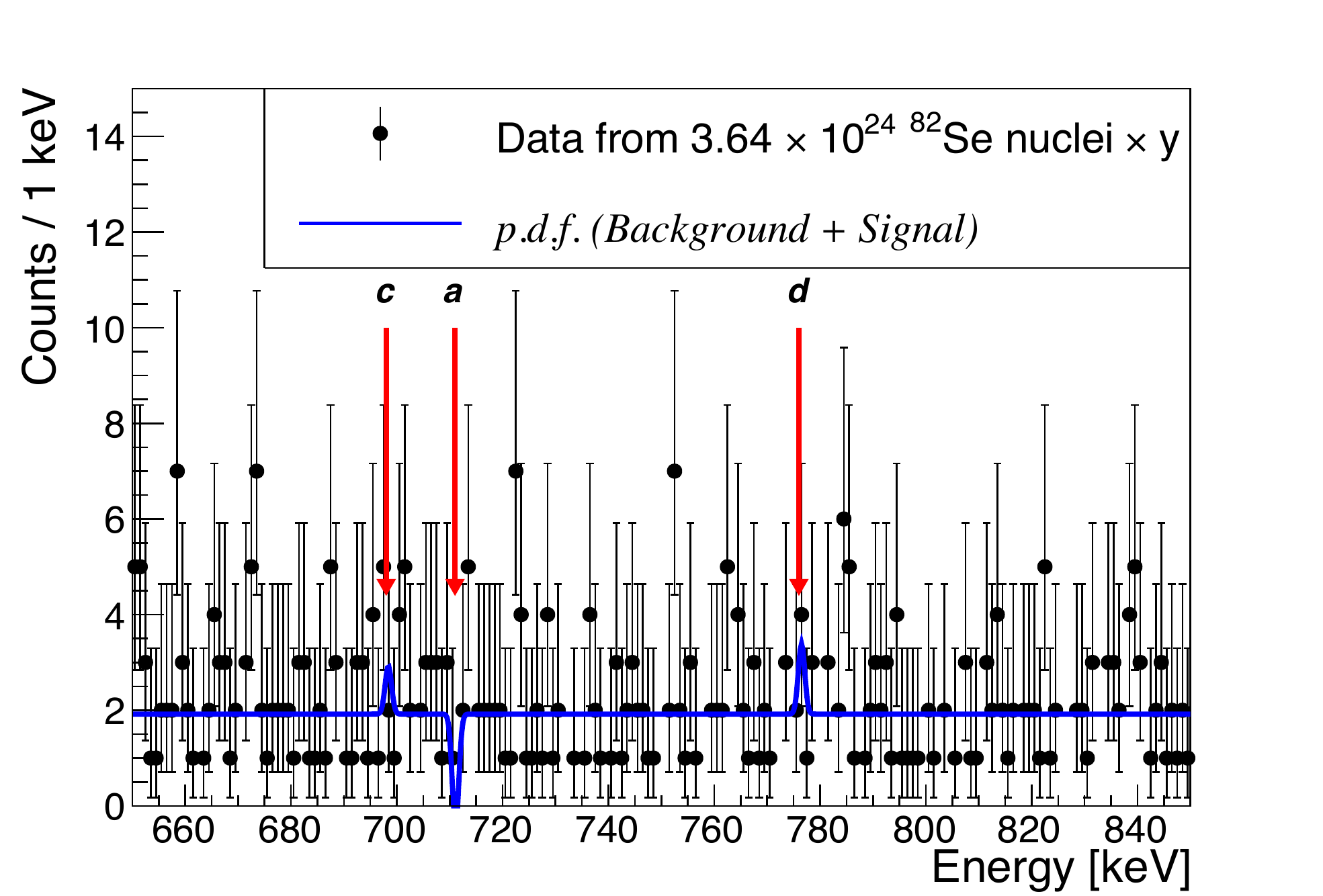}
\caption{\label{fig:Se82}Energy spectrum of $\beta\beta$ decay candidates in enriched $^{82}$Se (data points) and the best-fit model from the binned extended maximum likelihood analysis (solid blue line). The red arrows point at the three best fit Gaussian functions for the three regions of interest: $a$) for $^{82}\rm{Se} \rightarrow ^{82}\rm{Kr}_{0^{+}_{1}}$, $c$) $^{82}\rm{Se} \rightarrow ^{82}\rm{Kr}_{2^{+}_{1}}$ and $d$) for $^{82}\rm{Se} \rightarrow ^{82}\rm{Kr}_{2^{+}_{2}}$.}
\end{figure}

The fit result is shown in Fig.~\ref{fig:Se82}. Since no signal for any $\beta\beta$ decay of $^{82}$Se on exited levels is observed, a Bayesian lower limit at 90\% C.L. on the three decay rates, $\Gamma$s, is set. These limits are evaluated marginalizing the profile negative log likelihood function over nuisance parameters, assuming a flat prior on the $\Gamma$s when they assume physical values: $\pi(\Gamma)=1$ if $\Gamma \geq 0$, $\pi(\Gamma)=0$ otherwise. The limits evaluated for the decay rates are:
\begin{eqnarray}
&\Gamma_{0^{+}_{1}} < 2.0\times10^{-23} \rm{y^{-1}}\\
&\Gamma_{2^{2}_{1}} < 5.3\times10^{-23} \rm{y^{-1}}\\
&\Gamma_{2^{+}_{2}} < 6.7\times10^{-23} \rm{y^{-1}}.
\end{eqnarray}

The counting rate in the ROI is (9.6$\pm$0.5)~c/keV/y, which is 2~$\sigma$ compatible with the one measured in background conditions, (11.3$\pm$1.0)~c/keV/y,  when no $^{82}$Se sample was on the detector. Most likely, the background in the ROI is caused by detector internal contaminations~\cite{GeMPI}.

The corresponding lower limits on the partial half-lives, calculated as $T_{limit} = \ln(2) / \Gamma_{limit}$ are:
\begin{eqnarray}
&T^{\beta\beta}_{1/2} (^{82}\rm{Se} \rightarrow ^{82}\rm{Kr}_{0^{+}_{1}}) > 3.4\times10^{22} \rm{y}\\
&T^{\beta\beta}_{1/2} (^{82}\rm{Se} \rightarrow ^{82}\rm{Kr}_{2^{+}_{1}}) > 1.3\times10^{22} \rm{y}\\
&T^{\beta\beta}_{1/2} (^{82}\rm{Se} \rightarrow ^{82}\rm{Kr}_{2^{+}_{2}}) > 1.0\times10^{22} \rm{y}
\end{eqnarray}
Since the systematic uncertainties on the efficiency and live-time are at the level of \permil~their contributions do not affects the computed values significantly. 

The evaluated lower limit on the half-life of $^{82}$Se double-beta decay on the excited levels of daughter nuclei are the most stringent limits up-to-date. Our sensitivity surpasses the one of other works reported in literature~\cite{Se82_exc,Nemo_Se82} by about one order of magnitude. Furthermore the limit set for the decay to 2$_1^+$ state contradicts the theoretical calculations reported in~\cite{Dhiman_DBD}.

\section{Conclusions}

The goal of next-generation $\beta\beta$ decay experiments is to explore the inverted hierarchy region of neutrino mass. In order to further enhance their sensitivity, future experiments should strive for a highly efficient background reduction, possibly to a zero level, and large mass $\beta\beta$ sources, at the tonne-scale. These must exhibit low levels of radioactive contaminations in order to suppress the background in the region of interest. Furthermore, if the bolometric technique is adopted, low concentrations of chemical impurities in the source materials are also needed, given the fact that the detector performances depends on the quality of the materials used for the detector manufacturing.

In this work we have investigated the purity of 96.3\% enriched $^{82}$Se metal for the production of scintillating bolometers for the LUCIFER project, which amounts to 15~kg.  Out of the total, 2.5~kg were analyzed by means of $\gamma$ spectroscopy and ICP-MS analyses for the evaluation of the internal radioactive and chemical impurities.

The exceptional purity of the source, which shows extremely low internal radioactive contamination, at the level of tens of $\mu$Bq/kg and high chemical purity with an intrinsic overall contamination of few hundreds of 10$^{-6}$~g/g, allowed us to investigate also $\beta\beta$ decay of $^{82}$Se. We established the most stringent limits on the half-life of $^{82}$Se $\beta\beta$ decay to the excited levels 0$^+_1$, 2$^+_1$ and 2$^+_2$ of $^{82}$Kr . Previous limits are improved by one order of magnitude for all the investigated transitions. The technique used in this work does not allow for a distinction between the 2$\nu$ and the 0$\nu$ modes, therefore our results are the overlap of the two processes.

Our results give the opportunity to improve the accuracy of Nuclear Matrix Elements calculations, so that experimental and theoretical evaluations may have a direct comparison. Moreover, as discussed in~\cite{Suhonen_short_range}, fine tuning of parameters relevant for the quasiparticle random-phase approximation (QRPA) model, such as the particle-particle strength interaction, $g_{pp}$ can be performed on intermediate states of $\beta\beta$ decays.

\begin{acknowledgements}
This work was made in the frame of the LUCIFER experiment, funded by the European Research Council under the European UnionÕs Seventh Framework Programme (FP7/2007Ð2013)/ERC Grant Agreement no. 247115. 
\end{acknowledgements}

\end{document}